\documentstyle[12pt,psfig]{article}
\begin{document}
\newcommand{\cd}{\makebox[0.08cm]{$\cdot$}}

\centerline{\bf Anomalous $J/\psi$ suppression and charmonium dissociation
cross sections}
\vskip 18pt
\centerline{Xiao-Ming Xu$^1$, Cheuk-Yin Wong$^1$, and T. Barnes$^{1,2}$}
\vskip 14pt
\centerline{$^1$Physics Division, Oak Ridge National Laboratory, Oak Ridge,
Tennessee 37831}
\centerline{and}
\centerline{$^2$Department of Physics and Astronomy, University of Tennessee,
Knoxville, Tennessee 37996}

\begin{abstract}
\baselineskip=28pt
\noindent
We study $J/\psi$ suppression in Pb+Pb collisions at CERN-SPS energies
in hadronic matter with energy- and 
temperature-dependent charmonium dissociation cross sections 
calculated in the quark-interchange model of Barnes and Swanson. 
We find that the variation of J/$\psi$ survival probability
from peripheral to central collisions can be explained 
as induced by hadronic matter absorption in central collisions.   

\end{abstract}
\leftline{PACS codes: 25.75.-q, 13.75.Gx, 12.39.-x, 25.80.Ls}
\leftline{Keywords: $J/\psi$ suppression, hadronic matter, charmonium 
dissociation cross sections}

\newpage
\leftline{\bf I. INTRODUCTION}
\vspace{0.5cm}
\baselineskip=28pt
Following  the suggestion of Matsui and Satz [1] to probe 
quark-gluon plasma (QGP) with heavy quarkonium,  
the NA50 Collaboration investigated $J/\psi$ suppression in high-energy
heavy-ion collisions.
Anomalous $J/\psi$ suppression has been observed in
Pb+Pb collisions at $158A$ GeV [2-3]. 
 It is of interest to investigate whether deconfined 
matter is created in these collisions.
Many different mechanisms have been proposed to explain the phenomenon. 
It has been suggested that the suppression was due to a new phase, the 
QGP [4],
a change of the equation of state due to the  QCD phase transition [5], the 
melting of charmonia in QGP [6], or a percolation  deconfinement [7].
It has also been suggested that the anomalous suppression comes from the
absorption by various comovers produced in the Pb+Pb collisions [8-14]. 
   
In addition to the general suppression over a large region of transverse
energy $E_T$, which is related to  the impact parameter $b$, 
the data for  $J/\psi$ suppression in the large transverse energy 
region with
$E_T > 100$ GeV are also of special interest. 
This region  corresponds to the
most central Pb+Pb collisions. 
The possible presence of a rapid drop of the ratio of $J/\psi$
to Drell-Yan production cross sections, $B(J/\psi)\sigma (J/\psi) 
/\sigma (DY)$,  has been interpreted in a comover model as a 
result of multiplicity fluctuation, subject to a large uncertainty 
of the measured  data [15].
It has also been suggested that  large $E_T$ fluctuations lead 
to increasing probability
 for deconfinement and therefore possible complete suppresssion of $J/\psi$
[16]. A definitive understanding on 
the  nature of $J/\psi$ suppression in  Pb+Pb 
collisions is still lacking.       
 
The absorption of $J/\psi$ by comovers depends on the  $J/\psi$
dissociation  cross sections in collision with hadrons. The $J/\psi$
dissociation cross sections in collision with 
hadrons have been considered previously in several theoretical studies, but 
the predicted cross sections show great variation at low energies, largely 
due to different assumptions regarding the dominant scattering mechanism.
Kharzeev and Satz  
and collaborators [17] employed the parton model and perturbative 
QCD ``short-distance'' approach of Bhanot and Peskin [18] and found
remarkably  small low-energy cross sections for collisions of $J/\psi$
with light hadrons. Matinyan and M$\ddot {\rm u}$ller [19], 
Haglin [20], Lin and Ko [21], Song and Lee [22], and
Navarra and collaborators [23] recently 
reported results for these dissociation cross sections in meson exchange
models. These references all use effective meson Lagrangians, but differ in 
the interaction terms included in the Lagrangian. 
 Results depend on a choice
of meson coupling schemes and assumed form factors. 
Charmonium dissociation 
by nucleons has also been considered recently using  similar effective 
Lagrangian formulations [24].  

We favor the use of the known quark forces to obtain
the underlying scattering amplitudes from explicit nonrelativistic quark
model wavefunctions of the initial and final mesons [25, 26]. 
In this approach we first look 
for the inter-quark interaction by using the meson spectrum. The wave function
and  the interaction are then used to evaluate the cross section for the 
quark-interchange process [25, 26]. 
The calculated cross sections are first compared with experimental data of
$I=2$ $\pi \pi$ phase shifts, and excellent agreement between theory and
experimental data was obtained [25, 26].
Such an approach allows us to calculate all charmonium
dissociation cross sections by collisions with hadrons.  
Martins, Blaschke, and 
Quack previously reported dissociation cross section calculations using 
essentially the same approach [27].  
Two general 
characteristics of the cross sections 
 obtained in this approach are: (1) for endothermic reaction,
the cross section rises after the hadron energy exceeds some
threshold energies and the maximum magnitude of the cross section is in the
mb range, (2) for exothermic
reaction, the cross section is infinite  at zero kinetic
energy and decreases as the energy
increases.
      
The $J/\psi$ dissociation cross section depends on the meson wave functions and
the inter-quark interaction.  The latter quantities depend on temperature.
By using such a temperature-dependent inter-quark potential
inferred from lattice gauge calculations [28], the 
temperature dependence of $J/\psi$ dissociation cross sections can be 
obtained. 
The new cross sections are now applied to study
$B(J/\psi) \sigma (J/\psi)/\sigma (DY)$ versus $E_T$, to be 
compared with the CERN-SPS Pb+Pb data of the NA50 Collaboration. 

    The concept of the temperature dependence of the dissociation
process should be clarified as many different processes are involved.
There are two types of thermalization [26,29].  First, there is
the thermalization of the hadronic matter, in which the $J/\psi$ (or
its precursor) resides.  It occurs rapidly as it depends on the cross
section for the scattering of light hadrons with light hadrons, which
is of the order of tens of mb.  We expect that the dense hadronic
matter is thermalized within a time of the order of 1 fm/c after light
hadrons are produced.  (In our numerical results below, the
thermalization time for the hadronic matter is 1.9 fm/c after the
light hadrons are produced.)
 
    The second type of thermalization is the thermalization of the
``charmonium system" with its surrounding medium.  If a quarkonium is
placed in a thermalized hadronic medium, there will be
non-dissociative inelastic reactions between the quarkonium and medium
particles which change a charmonium state into another charmonium
state: $h + (Q\bar Q)_{JLS} \leftrightarrow h' + (Q\bar Q)_{J'L'S'}$.
These reactions lead to the thermalization of the charmonium system
[26,30].  When the heavy quarkonium system is in thermal
equilibrium with the medium, the probabilities for the occurrence of
heavy quarkonium states will be distributed according to the
Bose-Einstein distribution.  The time for the thermalization of such a
system depends on the non-dissociative inelastic cross sections.  As
these cross sections are different from the pion-pion scattering cross
section, the ``charmonium system" thermalization time is different
from the light-hadron matter thermalization time.

     The cross sections for non-dissociative inelastic reactions
between a charmonium and a light hadron and their temperature dependences
remain a subject of current
research.  Previous estimates of $\pi+J/\psi \to \pi + \psi'$ by Fujii
and Kharzeev [31] gave a cross section of the order of 0.01 mb.  In
the quark-interchange model of Barnes and Swanson [32,33], such a
process is forbidden in the first order and takes place only in the
second order. The higher order process presumably will lead to a smaller  
cross section. The non-dissociative inelastic cross sections
need to be evaluated for the interaction of pions with different charmonia
using different models, to provide an estimate of the time for
the thermalization of the charmonium system with the hadronic
medium. If the non-dissociative inelastic charmonium cross sections are of the 
same order as in $\pi - \pi$ collision, then the charmonium system will be also
thermalized and the dissociation of $J/\psi$ should be treated by the method
of statistical mechanics. On the other hand, 
if the non-dissociative inelastic cross sections are all of
the order of 0.01 mb, as estimated by Fujii and Kharzeev for
$\pi+J/\psi \to \pi + \psi'$, the thermalization time for the
charmonium system will be quite long and we will not need to consider
a thermalized charmonium system.  In the present work, we have  not
taken the process of dissociation by thermalization into account.

      For a charmonium placed in a thermal medium, the thermalized
hadronic matter alters the interaction between the charm quark and
charm antiquark and changes the QCD vacuum surrounding the charmonium.
The threshold energy for $J/\psi$ dissociation is shifted and
consequently the dissociation cross section changes as a function of
the hadronic matter temperature.  The dissociation cross section is
however of the order of a few mb [26].  Thus, in the treatment of the
collision of $J/\psi$ with the dense hadronic matter, the cross
section is still small enough so that we can treat the $\pi - J/\psi$
 dissociative  collision as a two-body process.

In Sec. II we describe
the inter-quark potential used to obtain meson wave functions.
In Sec. III we present the $\pi-J/\psi$ and $\pi-\chi_{cJ}$
dissociation cross sections 
and  show their dependences on temperature.  
In Sec. IV charmonium suppression in Pb+Pb collisions is examined.  
Numerical calculations and
results are shown and discussed in Sec. V. 
Conclusions are given in Sec. VI.

\vspace{0.5cm}
\leftline{\bf II. TEMPERATURE-DEPENDENT POTENTIAL AND WAVE FUNCTIONS}
\vspace{0.5cm}
In a medium at high temperatures, the gluon and light quark
fields fluctuate and the alignment of the color electric fields due to
the QCD interaction is reduced by the thermal motion for a random
orientation of the color electric fields. 
The ``pressure'' from the QCD vacuum to confine the quark and antiquark pair 
also diminishes with increasing temperature.
The inter-quark potential is thus a sensitive function of temperature,  
resulting in the vanishing of the string tension at the critical temperature
of deconfinement phase transition.

Even at $T=0$, a proper description of the heavy quarkonium state
should be based on a screening potential,
as the heavy quarkonium becomes a pair
of open charm or open bottom mesons when $r$ becomes very large, due
to the action of dynamical quark pairs [34].  Recently,
the  central   inter-quark potential 
has been obtained from the lattice gauge results of Karsch $et~al$. [28]
and analyzed by Digal $et~al$. [35, 36] and by Wong [26].
The temperature-dependent potential for $T<T_c$ can be
represented by
\begin{equation}
V_{12}(r,T)=- \frac {4}{3} \frac {\alpha_s e^{-\mu (T) r}}{r}
-\frac {b(T)}{\mu (T)} e^{-\mu (T) r},
\end{equation}
where $\alpha_s$ is a running coupling constant [25].
The effective string-tension coefficient is
\begin{equation}
b(T) = b_0 [1-(T/T_c)^2]\theta (T_c -T),
\end{equation}
with $b_0=0.35$ ${\rm GeV}^2$ and the effective screening parameter
\begin{equation}
\mu (T) = \mu_0 \theta (T_c -T),
\end{equation}
where $\mu_0 =0.28$ GeV and the step 
function $\theta$ signifies the vanishing of the string tension at the
phase transition temperature  $T_c =0.175$ GeV. 

The  charmonium  wave functions  can be obtained by solving
the Schr$\rm \ddot o$dinger equation  
\begin{equation}
[-\nabla \cdot \frac {1}{2\mu_{12}} \nabla +V_{12}(r,T)
+\Delta (r,T) ] \psi_{JLS} (\vec {r},T) = \epsilon (T) \psi_{JLS} (\vec {r},T),
\end{equation}
where $\mu_{12}$ is the reduced mass and  the mass difference 
$\Delta (r,T)$ is 
\begin{equation}
\Delta (r,T) =m_1 (r,T) +m_2 (r,T) -m_1 (\infty,T) -m_2 (\infty,T).
\end{equation}
The energy $\epsilon (T)$ is measured relative to two 
separated mesons $c\bar q$ and $q\bar c$ at $r \to \infty$
where  $c$ and $\bar c$ become open charm mesons  and 
$\{ m_1 (r,T), m_2 (r,T) \} =\{ M_{c\bar {q}}(T), M_{q\bar {c}}(T) \}$. 

The bound-state wave function   and energy $\epsilon$ have been obtained by 
diagonalizing the Hamiltonian using a nonorthogonal
Gaussian basis with different widths [26]. 
When the energy $\epsilon$ 
becomes positive, spontaneous dissociation occurs,
  subject to the selection
rules based on the conservation of total angular momentum and parity [26]. 

\vspace{0.5cm}
\leftline{\bf III. CHARMONIUM DISSOCIATION CROSS SECTIONS}
\vspace{0.5cm}
For meson-meson scatterings $A + B \to C +D$,
the differential cross section is given by
\begin{equation}
\frac {d\sigma_{fi}}{dt} =\frac {1}{64\pi s \mid \vec {p}_A \mid^2}
\mid {\cal M}_{fi} \mid^2,
\end{equation}
where $t$ is the momentum transfer squared, $s$ is the center-of-mass
(C.M.)  energy squared, and $\vec {p}_A$ is the momentum of 
meson $A$ in the C.M. system.  The matrix element
${\cal M}_{fi}$ is calculated with the four meson wave 
functions of $A$, $B$, $C$, and $D$ and the inter-quark interaction. The 
latter is the potential in Eq. (1) plus a spin-spin
 term for quark-quark interaction,  
\begin{equation}
V_{{\rm spin}-{\rm spin}} (r) =- \frac {\vec {\lambda}_1}{2} \cdot
\frac {\vec {\lambda}_2} {2} \times 
\frac {8\pi \alpha_s}{3m_1m_2} \vec {s}_1 \cdot \vec {s}_2 
(\frac {d^3}{\pi^{3/2}})e^{-d^2r^2}, 
\end{equation}
where $m_1 (m_2)$ and $\vec {s}_1 (\vec {s}_2)$ are individually the mass and
spin of quark 1(2), and 
$d$ is a spin-spin interaction width including relativistic corrections
[26, 37]. For quark-antiquark potential, $\frac {\vec {\lambda}_1}{2} \cdot
\frac {\vec {\lambda}_2} {2}$ is replaced by 
 $-\frac {\vec {\lambda}_1}{2} \cdot \frac {\vec {\lambda}_2^T} {2}$.
This potential generates wave functions which are then used to 
calculate the dissociation cross sections, as discussed in Ref. [26].
The dissociation          
cross section of $J/\psi$ in collision with $\pi$ varies as a function of 
$T/T_c$ and the C.M. kinetic energy $E_{KE} = \sqrt{s}-m_{A}-m_{B}$ where $m_{A}$ and $m_{B}$
are the masses of the incident mesons. We  plot in Fig. 1 
the $\pi +J/\psi$ dissociation cross sections.
Since about 35\% of $J/\psi$ comes from the
radiative decay of $\chi_{cJ}$, 
the dissociation cross sections of $\chi_{c1}$ and
 $\chi_{c2}$ in collisions with $\pi$ are also important. We evaluate these
cross sections which are plotted in Fig. 2. 
    
As the temperature increases, the temperature-dependent 
inter-quark potential becomes
weaker.  Then the energy for separating $c$ and $\bar{c}$ in the charmonium decreases and the
root-mean-square radius of the meson increases. As a consequence, 
 the dissociation cross 
section for a charmonium in collision with a pion rises with increasing temperatures. 
For various charmonia, different sizes also lead to
different dissociation cross sections.  
At $T/T_c \approx 0.7$, the  peak cross section of $\pi$ + $\chi_{c1}$
or $\pi$ + $\chi_{c2}$ is about 2.1
times that at $T=0$. Since the
$\chi_{c2}$  is less bound compared to $\chi_{c1}$, the dissociation
cross section for $\chi_{c2}$ is slightly larger. 
The $\pi +\chi_{c2}$ reaction
becomes exothermic at a lower temperature than $\pi + \chi_{c1}$.
Comparing Figs. 1 and 2, we find that the peak value of $\pi-J/\psi$
dissociation cross section is larger than  the peak values of 
$\pi-\chi_{c2}$ and $\pi-\chi_{c1}$ dissociation cross sections
when  $T/T_c$ are larger than 0.65  and 0.55, respectively.  

\vspace{0.5cm}
\leftline{\bf IV. CHARMONIUM SUPPRESSION IN Pb+Pb COLLISIONS}
\vspace{0.5cm}
Working in the nucleon-nucleon C.M. frame, 
we assume the following scenarios to investigate $J/\psi$ suppression. 
In a nucleus-nucleus collision,  hadronic matter  consisting mainly
of pions is produced at $\tau_{\rm form}$ after the collision. 
The hadronic matter is initially not in thermal equilibrium. 
It becomes thermalized by elastic
scatterings of pions at time $\tau_{\rm therm}$,
and the expansion of hadronic   
matter lowers the temperature and the  density until freeze-out at 
$\tau_{\rm fz}$. On the other hand,  
 $c\bar c$ pairs are created by the hard scattering processes between a 
projectile nucleon and a target nucleon 
in a very short time. 
The $c\bar c$ pair evolves into a charmonium or its precursor which
interacts and becomes dissociated by colliding 
first with nucleons in the two colliding nuclei and 
subsequently with  the produced hadronic matter.
$J/\psi$ will be dissociated by collision with pions during the time from
 $\tau_{\rm form}$ to $\tau_{\rm fz}$.    
The differential cross section for $J/\psi$ production
with respect to the impact parameter $\vec b$ is
\begin{eqnarray}
\frac {d\sigma^{AB}_{J/\psi}(\vec {b}) }{d\vec {b}} & = &
\int \frac {d^3p}{E} ( E\frac {d^3 \sigma^{NN}_{J/\psi}}{d^3p} )
\int \frac {d\vec {b}_A}{\sigma_{NJ/\psi}^2}
\{ 1-[1-T_A (\vec {b}_A) \sigma_{NJ/\psi} ]^A \} 
\{ 1-[1-T_B (\vec {b} -\vec {b}_A) \sigma_{NJ/\psi} ]^B \} \nonumber \\
& & \times \exp [-\int^{\tau_{\rm therm}}_{\tau_{\rm form}} d\tau v_{\rm rel} 
\sigma_{\pi J/\psi}  n_\pi (b,\tau) ]   \nonumber   \\
& & \times \exp [-\int^{\tau_{\rm {fz}}}_{\tau_{\rm therm}} d\tau 
\langle  v_{\rm rel} \sigma_{\pi J/\psi} \rangle n_\pi (b,\tau) ].
\end{eqnarray}

For the production of Drell-Yan dilepton pairs, the
cross section scales with the number of nucleon-nucleon collisions, and we have
\begin{equation}
\frac {d\sigma^{AB}_{DY}(\vec {b})}{d\vec {b}} = 
\sigma_{DY}^{NN} ABT_{AB} (\vec {b}). 
\end{equation}
The quantities $E$ and $\vec p$ are the energy and momentum of  
$J/\psi$; $\sigma_{J/\psi}^{NN}$ and $\sigma_{DY}^{NN}$ are 
cross sections for $J/\psi$ and Drell-Yan productions in a nucleon-nucleon 
collision, respectively; $\sigma_{\pi J/\psi}$ is the $J/\psi$ dissociation cross section in
collision with $\pi$; $\sigma_{NJ/\psi}$ is the $J/\psi$ absorption cross section in
collision with nucleons;  
$\vec {b}_A$ is the nucleon coordinate in the target nucleus $A$; 
$T_A$ and $T_B$ are nuclear thickness functions [38]; 
and $v_{\rm rel}$ is the relative velocity of charmonium and $\pi$. 
The thickness function for the colliding
 nuclei  is
\begin{equation}
T_{AB}(\vec {b})=\int d\vec {b}_A T_A(\vec {b}_A) T_B(\vec {b} -\vec {b}_A).
\end{equation}
The expression for the differential cross section for $\chi_{cJ}$ and $\psi'$
production can be obtained by replacing $J/\psi$ with $\chi_{cJ}$ or $\psi'$. 

If the hadronic matter is in thermal equilibrium, the pion number 
multiplicities are  
obtained from the familiar momentum distribution. 
Before thermalization, the pion number density has to be estimated 
 from experimentally measured multiplicities of charged pions. 
The total pion number   
$N_{\pi}$ as a function of the 
impact parameter $\vec b$ for Pb+Pb collisions at 158$A$ GeV
was measured by the NA49
collaboration [39]. We assume that the total pion number is conserved in the  
evolution of hadronic matter. The volume of hadronic matter depends on the
impact parameter $\vec b$ as well as the 
proper time $\tau$, which is measured 
relative to  the initial collision,
\begin{eqnarray}
V(b,\tau) & = &
\left ( 2\tau + \frac {\sqrt {4R_A^2-b^2}}{\gamma} \right )  \nonumber \\
& & \times \bigg \{ 2(R_A+\tau V_{{\rm ex}\bot})^2
\cos^{-1}(\frac {b}{2(R_A+\tau V_{{\rm ex}\bot})})
-b\sqrt {(R_A+\tau V_{{\rm ex}\bot})^2 -\frac {b^2}{4}} \bigg \}.       
\end{eqnarray}
Here $R_A$ is the lead-nucleus radius, $V_{{\rm ex}\bot}$ is the 
transverse velocity for radial flow, and $\gamma$ is the Lorentz contraction 
factor, $\gamma = \sqrt {s} /2m_N$, with nucleon mass $m_N$, and 
$\sqrt s$ is the 
C.M. energy of a nucleon-nucleon collision.  
As a result of the above consideration, the pion number density depends on the
impact parameter and proper time as
\begin{equation}
n_{\pi} (b,\tau )=\frac {N_\pi (b)}{V(b, \tau )}. 
\end{equation}

The 
$\pi$-charmonium dissociation cross sections depend on the temperature $T$ and
the collision energy. Pions in thermal equilibrium have the 
momentum distribution $f_\pi (k) \sim 3 e^{-E_\pi /T}$ with the pion energy 
$E_\pi$.
Integrating the whole pion momentum contribution to  $v_{\rm rel}\sigma $
and dividing by the pion number density, we obtain 
\begin{equation}
\langle v_{\rm rel} \sigma \rangle = 
\frac {\int \frac {d^3 k}{(2\pi)^3} f_\pi (k) v_{\rm rel} \sigma }
{\int \frac {d^3k}{(2\pi)^3} f_\pi (k) },
\end{equation}
which depends on $p_T$ and $x_F$ of charmonium and $\tau$.

The differential cross sections with respect to $E_T$ for
 $J/\psi$ production is given by
\begin{equation}
\frac {d\sigma_{J/\psi}^{AB}}{dE_T} =\int d^2b 
\frac {d\sigma_{J/\psi}^{AB}}{d\vec {b}} D(b,E_T),
\end{equation}
and the differential cross section for the Drell-Yan process is
\begin{equation}
\frac {d\sigma_{DY}^{AB}}{dE_T} =\int d^2b 
\frac {d\sigma_{DY}^{AB}}{d\vec {b}} D(b,E_T),
\end{equation}
where $D(b,E_T)$  relates  $b$ to $E_T$ [40]
\begin{equation}
D(b,E_T)=\frac {1}{\sqrt {2\pi}\sigma (b)}
e^{-[E_T -\bar {E}_T (b)]^2/2\sigma^2 (b)},
\end{equation}  
with the average transverse energy
\begin{equation}
\bar {E}_T (b) =\frac {135{\rm GeV}}{1+\exp [(b-6.7)/3] },
\end{equation}
and the variance
\begin{equation}
\sigma (b) = \frac {11{\rm GeV}}{1+\exp [(b-10)/2]}.
\end{equation}

\vspace{0.5cm}
\leftline{\bf V. RESULTS AND DISCUSSIONS}
\vspace{0.8cm}
\leftline{\bf A. Differential Cross Sections for $J/\psi$ Yield
in Nucleon-Nucleon Collisions}
\vspace{0.5cm}
To generate momentum distribution of $J/\psi$, we need to parametrize the
initially produced $J/\psi$ distributions.
The invariant differential cross section for $J/\psi$ production
in a nucleon-nucleon collision,
$Ed^3 \sigma^{NN}_{J/\psi}/d^3p$, is factorized as  
\begin{equation}
E\frac {d^3\sigma^{NN}_{J/\psi}}{d^3p}=f(x_F)g(p_T).
\end{equation}
The parameters of $f(x_F)$ and $g(p_T)$ obtained by fitting 
 experimental data in $p-p$ and $p-A$ 
collisions have been summarized in Refs. [41, 42]. We take 
the parametrizations
\begin{equation}
g(p_T) \sim (1+\frac {p_T^2}{\beta^2})^{-6},
\end{equation}
\begin{equation}
f(x_F) \sim \frac {(1-x_1)^a (1-x_2)^a}{x_1 + x_2},   
\end{equation}
where $x_{1,2} = 0.5(\sqrt {x_F^2 +4M_{J/\psi}^2/s} \pm x_F )$ with the
$J/\psi$ mass $M_{J/\psi}$. The quantity
$\beta$ is related to the average transverse
momentum $\langle p_T \rangle$ by $\beta = \frac {256}{35\pi} 
\langle p_T \rangle$. The quantity $\beta =2.24 \pm 0.28$
 GeV/$c$ is inferred from $\langle p_T \rangle =0.96 \pm 0.12$  GeV/$c$ 
for the $J/\psi$
production in $p$-Be collisions at $\sqrt {s} =16.8$ GeV [43]. 
The parameter $a = 4.95$ fits the $x_F$ spectra of 
$J/\psi$ in $p$-Be collisions at $\sqrt {s}=$
38.8, 31.6, and 16.8 GeV [43, 44] and $p$-Li collisions at 
$\sqrt {s}=23.8$ GeV [45]. The $x_F$ spectrum provides the average value
$\langle x_F \rangle  =\int_{0}^{1-\delta} dx_F x_F f(x_F) 
/\int_{0}^{1-\delta} dx_F f(x_F) \approx 0.16$ with
$\delta = M_{J/\psi}^{2}/s$.  The condition $x_{1,2} \leq$ 1 requires $x_{F} \leq$ 1-$\delta$.
We use the parametrizations  of  $f(x_F)$ and $g(p_T)$ to generate
 the $x_F$ and $p_T$ spectra of a charmonium
produced in nucleon-nucleon collisions in  Pb+Pb collisions at 
158 GeV/$c$ per nucleon.

In Eq. (8) we assume $J/\psi$ mesons are produced uniformly in the collision
region. The successive collisions of a projectile nucleon in the target 
nucleus cause the nucleon to lose energy gradually so that $J/\psi$ 
mesons are not evenly
produced in space. Such an effect was discussed in Ref. [46].

\vspace{0.5cm}
\leftline{\bf B. $J/\psi$ Suppression in Hadronic Matter}
\vspace{0.5cm}
As seen in Eq. (8), the differential cross section 
explicitly depends on the pion formation time $\tau_{\rm form}$, 
the pion thermalization time
$\tau_{\rm therm}$, and the hadron freeze-out time $\tau_{\rm fz}$. 
We estimate the formation
time $\tau_{\rm form}$ to be 1 fm/$c$ from  nucleon-nucleon collision 
data [38].
Then in the most central Pb+Pb collision, $n_{\pi}(b=0,\tau_{\rm form})= 
2.38$ ${\rm fm}^{-3}$. 
We use a freeze-out number density of $n_{\rm fz}=0.5$ ${\rm fm}^{-3}$.    
The initial temperature of pion matter is close to the
QCD phase transition  temperature $T_c =0.175$ GeV [28] and is taken as
$T_{\rm therm}=0.16$ GeV.

In a thermal medium, the charmonium will spontaneously  dissociate if the
 energy $\epsilon$ obtained from the Schr$\rm \ddot o$dinger equation
relative to the mass of two open charm mesons is positive. 
The  energy $\epsilon$ varies with temperature.
The critical temperature $T_d$ for the spontaneous dissociation (at
which $\epsilon$ is zero) was determined
by Wong [26] to be
 $T_d/T_c =0.99$, 0.90, and 
0.91 for $J/\psi$, $\chi_{c1}$, and $\chi_{c2}$, respectively.  Since 
$T/T_{c} \leq 
T_{\rm therm}/T_c =$ 0.91 for any centrality, we can neglect
the spontaneous
dissociation of $J/\psi$, $\chi_{c1}$, and $\chi_{c2}$  in the
pion matter.  

We assume that all charmonia concerned have 
the same absorption cross section in collision with  nucleons, 
$\sigma_{NJ/\psi} =\sigma_{N\chi_{c1}} =\sigma_{N\chi_{c2}}
=\sigma_{N\psi'}=\sigma_{\rm Nabs}$. Since these cross sections have not been
fixed by $p-A$ reactions, we take a value of 4.2 mb.
The remaining adjustable parameter is 
$\tau_{\rm therm} (b=0)$
for the most central Pb+Pb collision.
The thermalization time has a centrality dependence, 
$\delta \tau_{\rm therm} \sim 
1/(\sigma_{\pi \pi}n_\pi (b,\tau_{\rm form}))$,
where $\sigma_{\pi \pi}$ is the $\pi \pi$ elastic scattering cross section
at zero temperature. Since the pion number density is given by Eq. (12),
the thermalization time after the formation of hadronic matter can be obtained 
by
\begin{equation}
\delta \tau_{\rm therm} (b) =\delta \tau_{\rm therm} (b=0)
\frac {n_\pi (b=0,\tau_{\rm form})}{n_\pi (b,\tau_{\rm form})}.
\end{equation}
The thermalization time measured relative to the moment of initial collision is then
$\tau_{\rm therm}(b)=\tau_{\rm form} +\delta 
\tau_{\rm therm} (b)$.

The value of the adjustable parameter $\tau_{\rm therm}(b=0)$
and the transverse velocity $V_{{\rm ex}\bot}$ are
 obtained by fitting the latest data
of $B(J/\psi)\sigma (J/\psi)/\sigma (DY)$ from the NA50 Collaboration [47].
 As shown in Fig. 3. the latest data in Ref. [47]
are slightly different from the earlier 
data [2,3] in the peripheral and central collisions.
Other quantities such as the
pion number density at thermalization $n_{\rm therm}$, 
proper time $\tau_{\rm fz}$, and temperature $T_{\rm fz}$ at 
freeze-out are then calculated and also tabulated in 
 Table I for the central collision at $b=0$ fm. The quantity  
$n_{\rm therm}=n_{\pi}(b=0,\tau_{\rm therm})$ is calculated with Eq. (12). 
The freeze-out time $\tau_{\rm fz}$ depends on the centrality. 
The freeze-out temperature $T_{\rm fz}$,  determined in the most central
collision,  is the same for  hadronic matter yielded at
any centrality. 
The temperature  between 
 $T_{\rm therm}$ and $T_{\rm fz}$
approximately obey the relation given by Bjorken [48].      
In the survival probability of total $J/\psi$, we include contributions 
at the level of
58\% for direct $J/\psi$, 20\% for $\chi_{c1}$, 
14.5\% for $\chi_{c2}$, and 7.5\% for $\psi'$ [41].

From central to peripheral collisions, pion number density decreases and
thermalization time increases according to Eq. (22). At $b=8.9$ fm  
the pion number density at $\tau_{\rm therm}$
is equal to the freeze-out number density and no thermalization is
attained. For $b>8.9$ fm, 
the zero-temperature charmonium dissociation cross sections
are relevant for $J/\psi$ suppression. 
Since the peak dissociation cross sections for $J/\psi$, $\chi_{c1}$, and
$\chi_{c2}$ in collision with $\pi$ at $T=0$ 
are about 1.1 mb, 1.6 mb, and 1.9 mb, respectively, the
suppression of $J/\psi$ due to the collision with $\pi$ is very small.  
Therefore, the  absorption by $N-J/\psi$ collisions 
completely dominates $J/\psi$ suppression at
$b>8.9$ fm. For $b<8.9$ fm, 
thermalized hadronic matter sets in with  $T_{\rm therm}=0.16~{\rm GeV}$
and freezes out at $T_{\rm fz} =0.139$ GeV. 
The dissociation cross
section jumps from 1.1 mb (1.6 mb, 
1.9 mb) at $T=0$  to 6.5 mb (4.1 mb, 4.8 mb) at 
$T_{\rm fz}=0.139$ GeV and even higher at $T_{\rm therm}=0.16~{\rm GeV}$ for 
$J/\psi$ ($\chi_{c1}$, $\chi_{c2}$). But this sudden rise does not induce a
sudden fall of $B(J/\psi) \sigma (J/\psi )/\sigma (DY)$ at
$E_T \approx 40$ GeV since the time in which $J/\psi$ interacts with pion 
matter in thermal equilibrium
increases gradually   as the impact parameter decreases from
$b=8.9$ fm to zero. 
In these calculations, the suppression of $\psi'$ has been
taken into account (see
the following paragraph), as $\psi'$ comprises about 7.5\% of the 
unsuppressed $J/\psi$ yield. 

To study $\psi'$ suppression, we note from  
Ref. [26] that the spontaneous dissociation of $\psi'$
begins with $T_d/T_c=0.50$. The pion matter freezes out at $T_{\rm fz}/T_c 
\approx 0.79$ for any centrality. Thus $\psi'$ spontaneous dissociation takes 
place throughout the pion matter. For a full description of  the 
$\psi'$ spontaneous dissociation, a suitable
treatment will be  needed  in future work. The effect of $\psi'$
dissociation,  alternatively,
can be accounted for by an approximate effective treatment where 
we assume a constant $\psi'$
dissociation cross section independent of energy, 
$\sigma_{\pi \psi'}=12$ mb. 
As shown in Fig. 4, the experimental
data for $B(\psi')\sigma(\psi')/\sigma(DY)$ reported in Ref. [49] 
and exhibited in Ref. [42] can be represented by such
a $\pi-\psi'$ cross section of 12 mb. 
We shall include this  constant $\psi'$ dissociation cross section  
in our analysis of the $\psi'$ absorption in our model. It can be
considered as an effective representation of the experimental data to include
the effect of $\psi'$ suppression.

In addition to the spontaneous dissociation, $J/\psi$ suppression may receive
 contributions from the thermalization of charmonium in hadronic matter. The 
 thermalization  process leads to a thermal distribution of charmonium, and the
states above the dissociate threshold can dissociate spontaneously [26, 30]. 
The thermalization of charmonia depends on the dynamic processes such as 
scatterings of lower charmonium states in collision with pions leading to 
higher charmonium states. 
The work of Fujii and Kharzeev [31] gave the cross
sections for $\pi + J/\psi \to \pi +\psi'$ less than 0.01 mb in  the region of
interest. We assume implicit generalization of this cross section to other 
pion-charmonium scatterings. 
Then the thermalization time of charmonia is very
 long and the  charmonia cannot be in thermal equilibrium
 before pionic matter freezes out. 
Therefore the dissociation by thermalization for
 $J/\psi$, $\chi_{c1}$, and $\chi_{c2}$ have not been included. 
If we calculate cross sections for processes such as $\pi + J/\psi \to \pi 
+\chi_{cJ}$  in the quark-interchange mechanism, we need to interchange the
quarks twice and the quark-interchange processes are of higher order. 
The calculations for these cross sections are complicated and are left 
for a future work.  

To illustrate the role of  charmonium dissociations in collisions with pions 
in hadronic matter, we set $\sigma_{\rm Nabs}$=0 and calculate 
$B(J/\psi)\sigma (J/\psi)/\sigma
(DY)$ with different assumptions on the  
$\pi - (c\bar {c})_{JLS}$ cross 
sections for the dissociation of the charmonium $(c\bar {c})_{JLS}$. 
In Fig. 5 we show the results with
$\pi- (c\bar {c})_{JLS}$ dissociation cross sections  as given by: 
(1) $\sigma (\pi-(c\bar {c})_{JLS})$ for $T=0$ [25] (dashed curve);
(2) $\sigma (\pi-(c\bar {c})_{JLS})$=1 mb (dot-dashed curve); 
(3) $\sigma (\pi-(c\bar {c})_{JLS})$=2 mb (dotted curve); 
(4) $\sigma (\pi-(c\bar {c})_{JLS})$ of Figs. 1 and 2 for $T>0$ (solid curve).
The flat curve below $E_T =28$ GeV comes mainly from the collisions at 
$b>$8.9 fm where no thermalization is attained. 
This flatness arises as
 the $\pi$-charmonium dissociation cross section is
small and the pion number density is not high at $b >$ 8.9 fm. 
Indeed, the dashed curve
gives a  small $J/\psi$ suppression if only the $\pi$-charmonium
dissociation cross sections at $T=0$  
are used. Even the $J/\psi$ suppression obtained by using
 1 mb cross section is larger than that for the cross sections at
$T=0$. It is obvious that the suppression obtained from the cross 
sections at higher temperatures 
is  greater than that from using a 2 mb cross section. 
This is related to the larger charmonium dissociation cross sections at 
higher temperatures as shown below.

The  cross sections in Figs. 1 and 2
depend on the temperature and $\pi -(c\bar {c})_{JLS}$ relative momenta. 
What are 
the average cross sections? Fig. 6 exhibits cross sections
averaged over the pion and $(c\bar {c})_{JLS}$ 
 momenta in the  collision at $b=0$ fm, which is defined as
\begin{equation}
\langle \sigma \rangle = 
 \int \frac {d^3p}{E} ( E\frac {d^3 \sigma^{NN}_{J/\psi}}{d^3p} )
\frac {d^3 k}{(2\pi)^3} f_{\pi} (k) \sigma  \bigg /
[ \int  \frac {d^3p}{E} ( E\frac {d^3 \sigma^{NN}_{J/\psi}}{d^3p} )
\int \frac {d^3 k}{(2\pi)^3} f_{\pi} (k) ].
\end{equation}
The average cross sections $\langle \sigma_{\pi J/\psi} \rangle$, 
$\langle \sigma_{\pi \chi_{c1}} \rangle$, and 
$\langle  \sigma_{\pi \chi_{c2}} \rangle$ decrease with time.
At the thermalization time, they are 
3.47 mb, 2.74 mb, and 2.95 mb, respectively.
At 2.9 fm$<\tau <$4.25 fm,  $\langle \sigma_{\pi J/\psi} \rangle$ is larger
than $\langle \sigma_{\pi \chi_{c1}} \rangle$. At 2.9 fm $<\tau <$3.4 fm,
 $\langle \sigma_{\pi J/\psi} \rangle$ even surpasses
 $\langle \sigma_{\pi \chi_{c2}} \rangle$. These are not surprising
since the peak value of $J/\psi$ dissociation cross section exceeds that 
of $\chi_{c1}$ ($\chi_{c2}$) for $T/T_c > 0.55(0.65)$  
and   the temperature of hadronic
matter before freeze-out is  $T/T_c \geq 0.79$. 

We further calculate the quantity
\begin{equation}
\overline {v_{\rm rel} \sigma} =\int^{\tau_{\rm fz}}_{\tau_{\rm therm}} d\tau
\langle v_{\rm rel} \sigma \rangle /(\tau_{\rm fz} -\tau_{\rm therm}). 
\end{equation}
The $p_T$ and $x_F$ dependences of $\overline {v_{\rm rel} \sigma}$ 
are shown in Figs. 7 and 8. 
These results indicate that the average absorption effects due to $J/\psi$, 
$\chi_{c1}$, and $\chi_{c2}$ are not significantly different.
The above discussion will be useful to understand data from 
the NA60 Collaboration [50] where the $\chi_{cJ}$ suppressions will be
measured.

\vspace{0.5cm}
\leftline{\bf C. $J/\psi$ Suppression at Large $E_T$}
\vspace{0.5cm}
In addition to the general suppression over a broad region of transverse
energy $E_T$, 
the data for  $J/\psi$ suppression in the large transverse energy 
region with $E_T > 100$ GeV are also of special interest. 
For $E_T>100$ GeV, we consider the effects 
arising from the large $E_T$
fluctuation. Since $D(b,E_T)$ has not been constrained to give $E_T$ by
the average
\begin{equation} 
\langle \bar {E}_T \rangle  
= \frac {\int d^2b \bar {E}_T(b) D(b,E_T)}{\int d^2b D(b,E_T)},
\end{equation}  
the measured $E_T$ is larger than $\langle \bar {E}_T \rangle$ for 
$E_T > 100$ GeV, 
i.e. the so-called large-$E_T$ fluctuation happens [15,16]. In our 
calculations, the fluctuation increases not only the pion number    
density by the replacement $n_{\pi} (b,\tau ) \to n_{\pi} (b,\tau )  
E_{T}/\langle \bar{E}_T \rangle$    
but also the lifetime of pion matter by $V(b,\tau_{\rm fz}(b))=
V(b, \tau_{\rm therm}(b)) n_{\pi} (b, \tau_{\rm therm}(b))/n_{\rm fz}$. 
The two changes lead to the solid curve in Fig. 3 
decreasing continually beyond $E_T =100$ GeV.

The large $E_T$ fluctuation does not increase the initial temperature since
we do not consider here the QCD phase transition to a QGP.  
The spontaneous dissociations  of $J/\psi$, $\chi_{c1}$, and $\chi_{c2}$ can
still be neglected.  
The factor $E_{T}/\langle \bar{E}_{T} \rangle$ 
shows the fluctuation increases smoothly from 
$E_T =100$ GeV. Therefore we see that the solid curve does not drop rapidly
in the large-$E_T$ region and does not 
pass the rightest experimental point within the error
bar. This behavior is similar to the theoretical $J/\psi$ suppression obtained
by Capella $et~al$. [15]. If  experimental data  
at $E_T>120$ GeV can be provided with good statistics
by additional NA60 measurements [50], 
then we need compare our result to the experimental data in this region  
to see the possibility of additional suppression due to the QGP.

\vspace{0.5cm}
\leftline{\bf VI. CONCLUSIONS}
\vspace{0.5cm}
We have studied $\pi$-charmonium dissociation cross sections and
the $J/\psi$ suppression in Pb+Pb collisions
at the CERN-SPS. We use dissociation cross sections calculated in the
Barnes-Swanson quark-interchange model [32,33] with the potential 
obtained from the lattice gauge results. 
Making reasonable assumptions about the absorption scenario, we find that
the $J/\psi$ production cross section calculated with these dissociation
cross sections can describe the general features
of the anomalous $J/\psi$ suppression data.
An important element of the agreement arises from the increase of the 
dissociation cross sections as the temperature increases. 
For the large $E_T$ region at $E_T >100$ GeV, the anomalous additional 
suppression can be explained by
the fluctuation of $E_T$ or multiplicity. 

The $\pi$-charmonium dissociation cross sections depend on the temperature and the kinetic
energy, but
the averages $\overline 
{v_{\rm rel}\sigma}$ in hadronic matter for
$\pi +J/\psi$, $\pi +\chi_{c1}$, and $\pi +\chi_{c2}$ 
dissociation cross sections are nearly the same. Thus, 
the dissociations of direct $J/\psi$, 
$\chi_{c1}$, and $\chi_{c2}$ 
give nearly the same contributions to the 
 suppression of the measured $J/\psi$.

We have chosen the constant absorption cross sections of charmonia on
 nucleons. However, the absorption cross sections are expected to depend on  
the C.M. energy of charmonium and nucleon. Determining the nucleon
absorption cross  sections is an important task. 
It  will be of interest to evaluate the $N-J/\psi$ dissociation cross sections
 in the Barnes-Swanson quark-interchange model [32,33].

\vspace{0.5cm}
\leftline{\bf ACKNOWLEDGEMENTS}
\vspace{0.5cm}
X.-M. Xu thanks the nuclear theory group and PHENIX group at ORNL for their
kind hospitality. 
This work was supported in part by the Division of Nuclear Physics, 
U. S. Department of Energy, and by the Laboratory Directed Research and
Development of Oak Ridge National Laboratory, 
under Contract No. DE-AC05-00OR22725 managed by 
UT-Battelle, LLC. X.-M. Xu acknowledges support from the CAS Knowledge
Innovation Project No. KJCX2-SW-N02.

\newpage
\centerline{\bf References}
\vskip 14pt
\leftline{[1]T. Matsui and H. Satz, Phys. Lett. B178, 416 (1986).}
\leftline{[2]NA50 Collaboration, M. C. Abreu et al.,
Phys. Lett. B450, 456 (1999).}
\leftline{[3]NA50 Collaboration, M. C. Abreu et al.,
Phys. Lett. B477, 28 (2000).}
\leftline{[4]J.-P. Blaizot and J.-Y. Ollitrault, Phys. Rev. Lett. 77, 1703 (1996).}
\leftline{~~~C. Y. Wong, Nucl. Phys. A610, 434c (1996); A630, 487c (1998).}
\leftline{~~~D. Kharzeev, C. Lourenco, M. Nardi, and H. Satz, Z. Phys. C74, 
307 (1997).}
\leftline{[5]E. Shuryak and D. Teaney, Phys. Lett. B430, 37 (1998).}
\leftline{[6]R. Vogt, Phys. Lett. B430, 15 (1998).}
\leftline{[7]M. Nardi and H. Satz, Phys. Lett. B442, 14 (1998).}
\leftline{[8]S. Gavin and R. Vogt, Nucl. Phys. A610, 442c (1996).}
\leftline{~~~W. Cassing and C. M. Ko, Phys. Lett. B396, 39 (1997).}
\leftline{~~~W. Cassing and E. L. Bratkovskaya, Nucl. Phys. A623, 570 (1997).}
\leftline{[9]N. Armesto and A. Capella, Phys. Lett. B430, 23 (1998).}
\leftline{[10]J. D. de Deus and J. Seixas, Phys. Lett. B430, 363 (1998).}
\leftline{[11]J. H$\rm \ddot u$fner and B. Z. Kopeliovich, Phys. Lett. B445, 
223 (1998).}
\leftline{[12]J. Geiss, C. Greiner, E. L. Bratkovskaya, W. Cassing, and 
U. Mosel, Phys. Lett.} 
\leftline{~~~~~B447, 31 (1999).}
\leftline{[13]D. E. Kahana and S. H. Kahana, Phys. Rev. C59, 1651 (1999).}
\leftline{[14]D. Prorok and L. Turko, Phys. Rev. C64, 044903 (2001).}
\leftline{[15]A. Capella, E. G. Ferreiro, and  A. B. Kaidalov,
Phys. Rev. Lett. 85, 2080 (2000).}
\leftline{[16]J.-P. Blaizot, P. M. Dinh, and J.-Y. Ollitrault, Phys.
Rev. Lett. 85, 4012 (2000).}
\leftline{[17]D. Kharzeev and H. Satz, Phys. Lett. B334, 155 (1994).}
\leftline{~~~~~D. Kharzeev, H. Satz, A. Syamtomov, and  G. Zinovjev, Phys. 
Lett. B389, 595 (1996).}
\leftline{[18]M. E. Peskin, Nucl. Phys. B156, 365 (1979).}
\leftline{~~~~~G. Bhanot and M. E. Peskin, Nucl. Phys. B156, 391 (1979).}
\leftline{[19]S. G. Matinyan and B. M$\rm \ddot u$ller, Phys. Rev. C58, 
2994 (1998).}
\leftline{[20]K. L. Haglin, Phys. Rev. C61, 031902 (2000).}
\leftline{~~~~~K. L. Haglin and C. Gale, Phys. Rev. C63, 065201 (2001).}  
\leftline{[21]Z. Lin and C. M. Ko, Phys. Rev. C62, 034903 (2000).}
\leftline{[22]Y. Oh, T. Song, and S. H. Lee, Phys. Rev. C63, 034901 (2001).}
\leftline{[23]F. S. Navarra, M. Nielsen, and M. R. Robilotta, Phys. Rev. C64,
021901 (2001).}
\leftline{[24]A. Sibirtsev, K. Tsushima, and A. W. Thomas, Phys. Rev. C63, 
044906 (2001).}
\leftline{~~~~W. Liu, C. M. Ko, and Z. W. Lin, Phys. Rev. C65, 015203 (2001).}
\leftline{[25]C. Y. Wong, E. S. Swanson, and T. Barnes, Phys. Rev. C62, 
045201 (2000);} 
\leftline{~~~~C65, 014903 (2001).}
\leftline{[26]C. Y. Wong, Phys. Rev. C65, 034902 (2002) [nucl-th/0110004].}
\leftline{[27]K. Martins, D. Blaschke, and E. Quack, Phys. Rev. C51, 2723 (1995).}
\leftline{[28]F. Karsch, E. Laermann, and A. Peikert, Nucl. Phys. B605, 
579 (2001).} 
\leftline{[29]C. Y. Wong, T. Barnes, E. S. Swanson, and H. W. Crater, 
talk presented at Quark} 
\leftline{~~~~Matter Conference 2002, July 18-24, 2002, Nante, France, to
appear in the Proceedings,}
\leftline{~~~~nucl-th/0209017.}
\leftline{[30]D. Kharzeev, L. McLerran, and H. Satz, Phys. Lett. B356, 349 (1995).}
\leftline{[31]H. Fujii and D. Kharzeev, Phys. Rev. D60, 114039 (1999).}
\leftline{[32]T. Barnes and E. S. Swanson, Phys. Rev. D46, 131 (1992).}
\leftline{[33]E. S. Swanson, Ann. Phys. (N.Y.)220, 73 (1992).} 
\leftline{[34]F. Karsch, M. T. Mehr, and H. Satz, Z. Phys. C37, 617 (1988).}  
\leftline{~~~~~C. Y. Wong, Phys. Rev. D60, 114025 (1999).}
\leftline{~~~~~K. J. Juge, J. Kuti, and C. Morningstar, hep-lat/9809015.} 
\leftline{[35]S. Digal, D. Petreczky, and H. Satz, Phys. Lett. B514, 57 (2001).}
\leftline{[36]S. Digal, D. Petreczky, and H. Satz, Phys. Rev. D64, 094015 (2001).}
\leftline{[37]S. Godfrey and N. Isgur, Phys. Rev. D32, 189 (1985).}
\leftline{[38]C. Y. Wong, {\it Introduction to High-Energy Heavy-Ion 
Collisions} (World Scientific,} 
\leftline{~~~~~Singapore, 1994).}  
\leftline{[39]NA49 Collaboration, F. Sikl$\rm {\acute e}$r et al., Nucl.
Phys. A661, 45c (1999).}
\leftline{[40]C. Y. Wong, ORNL-CTP-9805 and hep-ph/9809497.}
\leftline{[41]G. A. Schuler, CERN-TH/94-7170.}
\leftline{[42]R. Vogt, Phys. Rep. 310, 197 (1999).}
\leftline{[43]K. J. Anderson et al., Phys. Rev. Lett. 37, 799 (1976).}
\leftline{[44]E672 and E706 Collaborations, A. Gribushin et al., Phys. Rev.
D62, 012001 (2000).}
\leftline{[45]E705 Collaboration, L. Antoniazzi et al., Phys. Rev. D46, 
4828 (1992).}
\leftline{[46]C. Gale, S. Jeon, and J. Kapusta, Phys. Rev. C63, 024901 (2001).}
\leftline{[47]NA50 Collaboration, L. Ramello et al., in {\it Proceedings of
Quark Matter 2002}, Nantes,} 
\leftline{~~~~~France, July 18-24, 2002.} 
\leftline{[48]J. D. Bjorken, Phys. Rev. D27, 140 (1983).}
\leftline{[49]NA50 Collaboration, L. Ramello et al., 13th International 
Conference on}
\leftline{~~~~~Ultrarelativistic Nucleus--Nucleus Collisions, Tuskuba, Japan, 
1997.}
\leftline{[50]NA60 proposal, CERN/SPSC 2000-010, 7 March 2000;}
\leftline{~~~~~NA60 memorandum, CERN/SPSC 2001-009, 12 March, 2001;}
\leftline{~~~~~NA60 memorandum, CERN/SPSC 2002-009, 12 March, 2002.}

\newpage
\begin{table}
\caption{Values of $\sigma_{\rm Nabs}$, $V_{{\rm ex}\bot}$,
 $\tau_{\rm therm}$, $n_{\rm therm}$, 
$\tau_{\rm fz}$, and $T_{\rm fz}$.}
\begin{center}
\begin{tabular}{c|c|c|c|c|c}
\hline\hline
{\sl $\sigma_{\rm Nabs}({\rm mb})$}   
& {\sl $V_{{\rm ex}\bot}(c)$ }
& {\sl $\tau_{\rm therm}({\rm fm}/c)$}
& {\sl $n_{\rm therm}({\rm fm}^{-3})$} & {\sl $\tau_{\rm fz}({\rm fm}/c)$} 
& {\sl $T_{\rm fz}({\rm GeV})$}       \\
\hline
{4.2} & {0.75} & {2.9} & {0.88} & {4.44}  & {0.139}    \\
\hline\hline
\end{tabular}
\end{center}
\end{table}

\newpage
\begin{figure}[t]
  \begin{center}
  \end{center}
\label{fig0}
\end{figure}

\newpage
\begin{figure}[t]
  \begin{center}
    \leavevmode
    \parbox{\textwidth}
           {\psfig{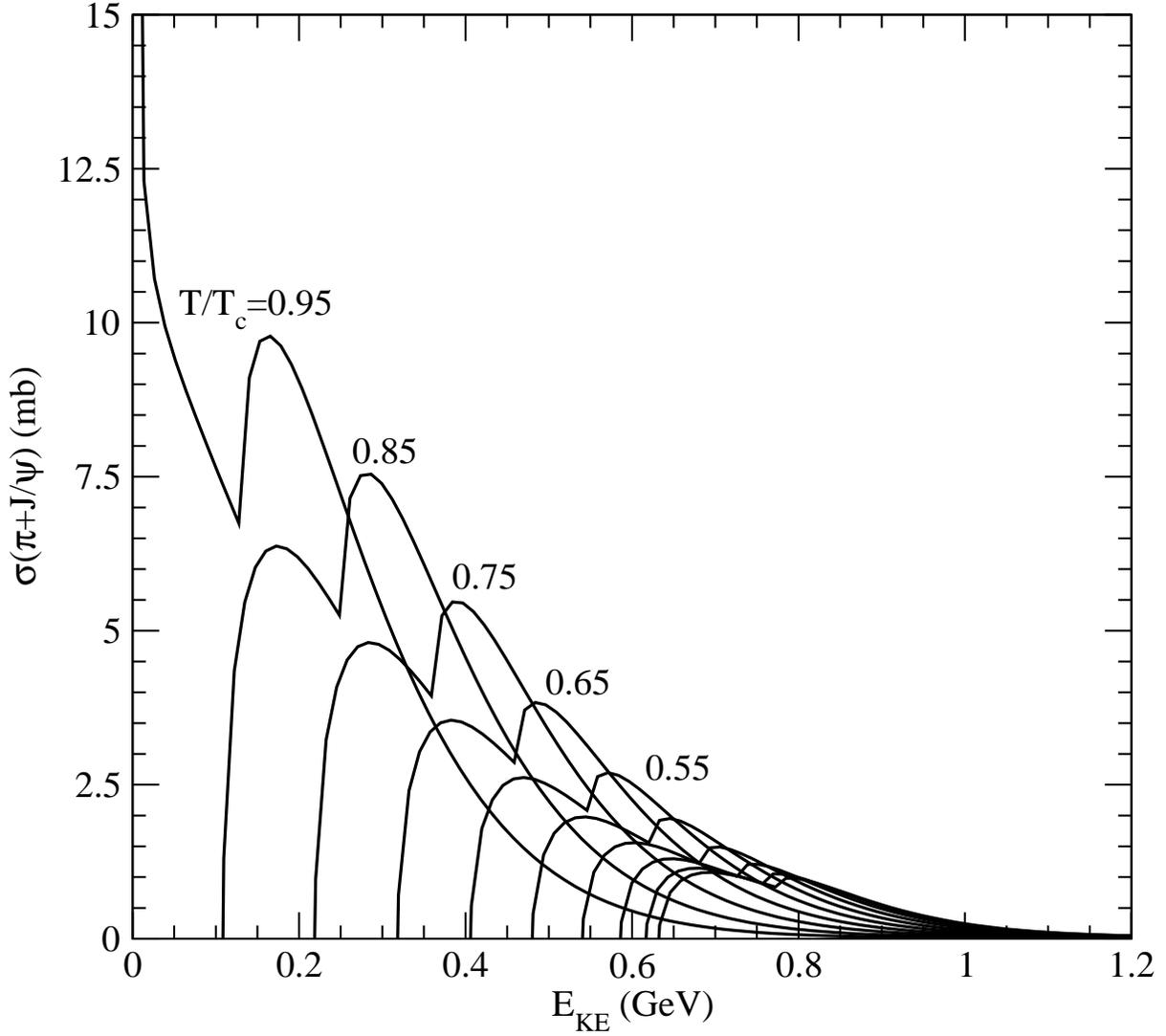}}
  \end{center}
\caption{
$J/\psi$ dissociation cross sections for various temperatures as a
function of $E_{KE}$. From left to right are
$T/T_c$=0.45, 0.35, 0.25, 0.15, 0.05 for these curves without labels.
}
\label{fig1}
\end{figure}

\newpage
\begin{figure}[t]
  \begin{center}
    \leavevmode
    \parbox{\textwidth}
           {\psfig{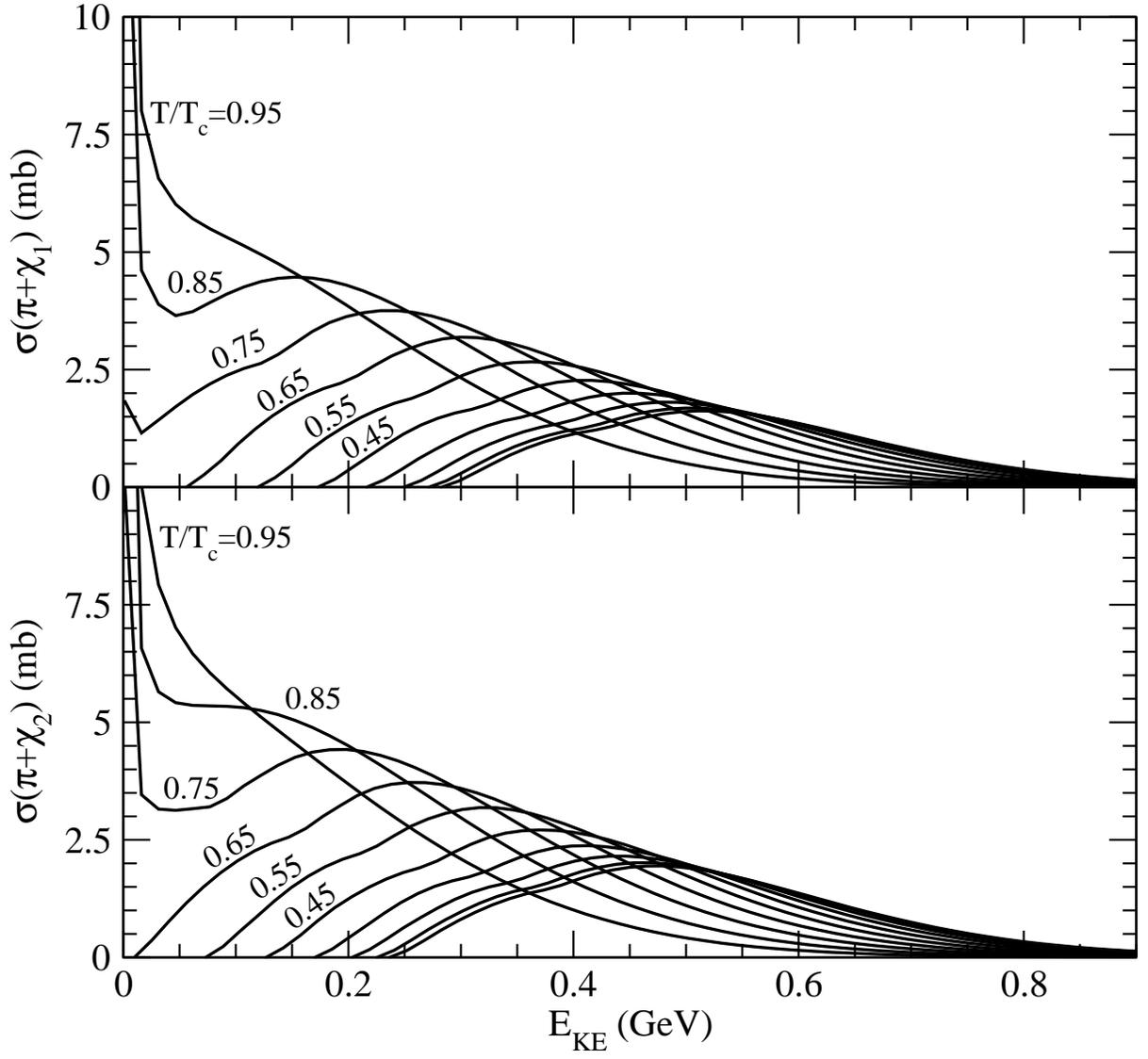}}
  \end{center}
\caption{
Upper and lower panels show
$\chi_{c1}$ and $\chi_{c2}$ dissociation cross sections 
for various temperatures as a function of $E_{KE}$. From left to right are
$T/T_c$=0.35, 0.25, 0.15, 0.05 for these curves without labels.
}
\label{fig2}
\end{figure}

\newpage
\begin{figure}[t]
  \begin{center}
    \leavevmode
    \parbox{\textwidth}
           {\psfig{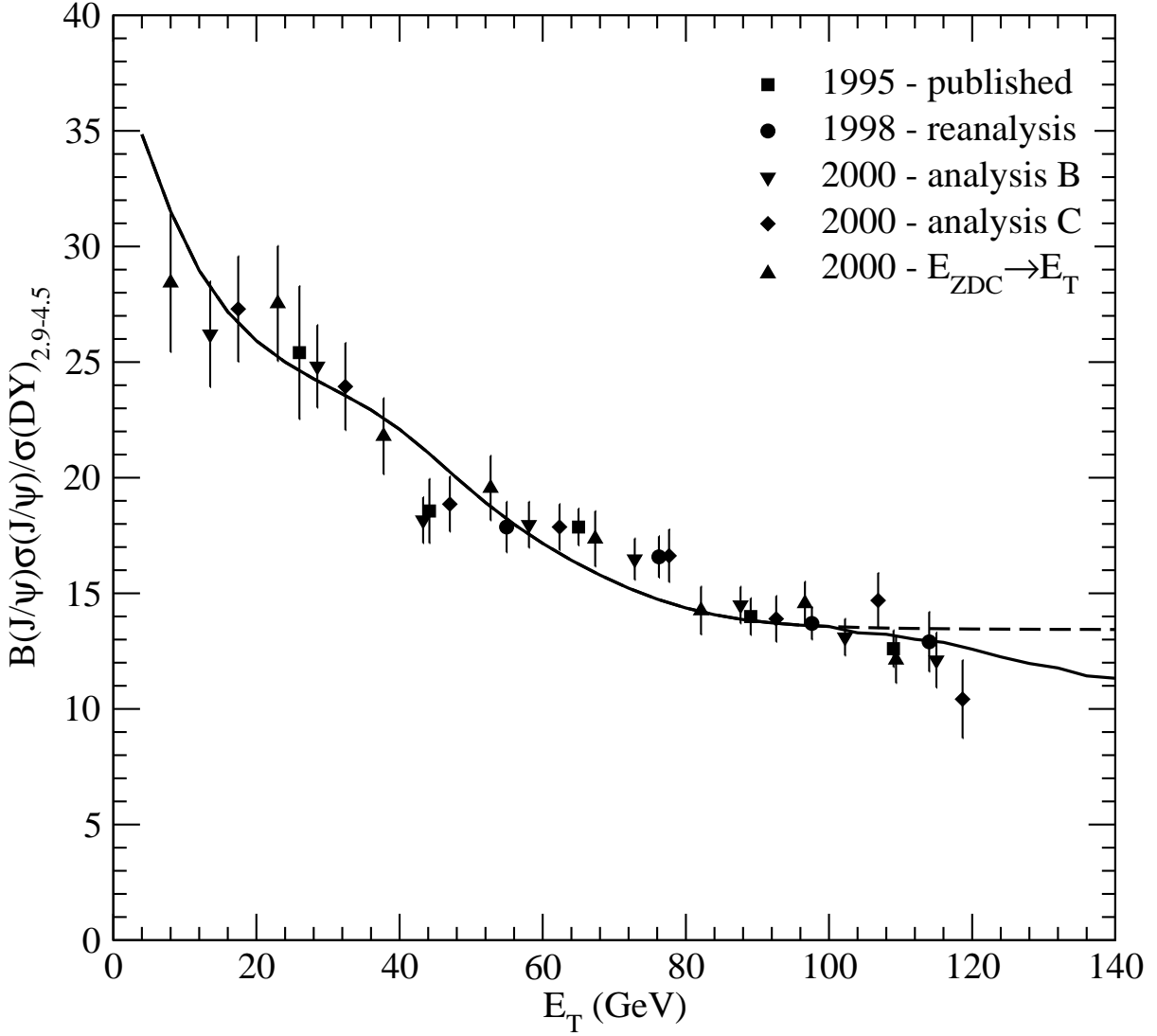}}
  \end{center}
\caption{
The experimental data are from Ref. [47]. 
The solid curve is
the theoretical result while $\sigma_{\rm Nabs} =4.2$ mb.
The dashed curve is 
obtained with no large-$E_T$ fluctuation considered.
}
\label{fig3}
\end{figure}

\newpage
\begin{figure}[t]
  \begin{center}
    \leavevmode
    \parbox{\textwidth}
           {\psfig{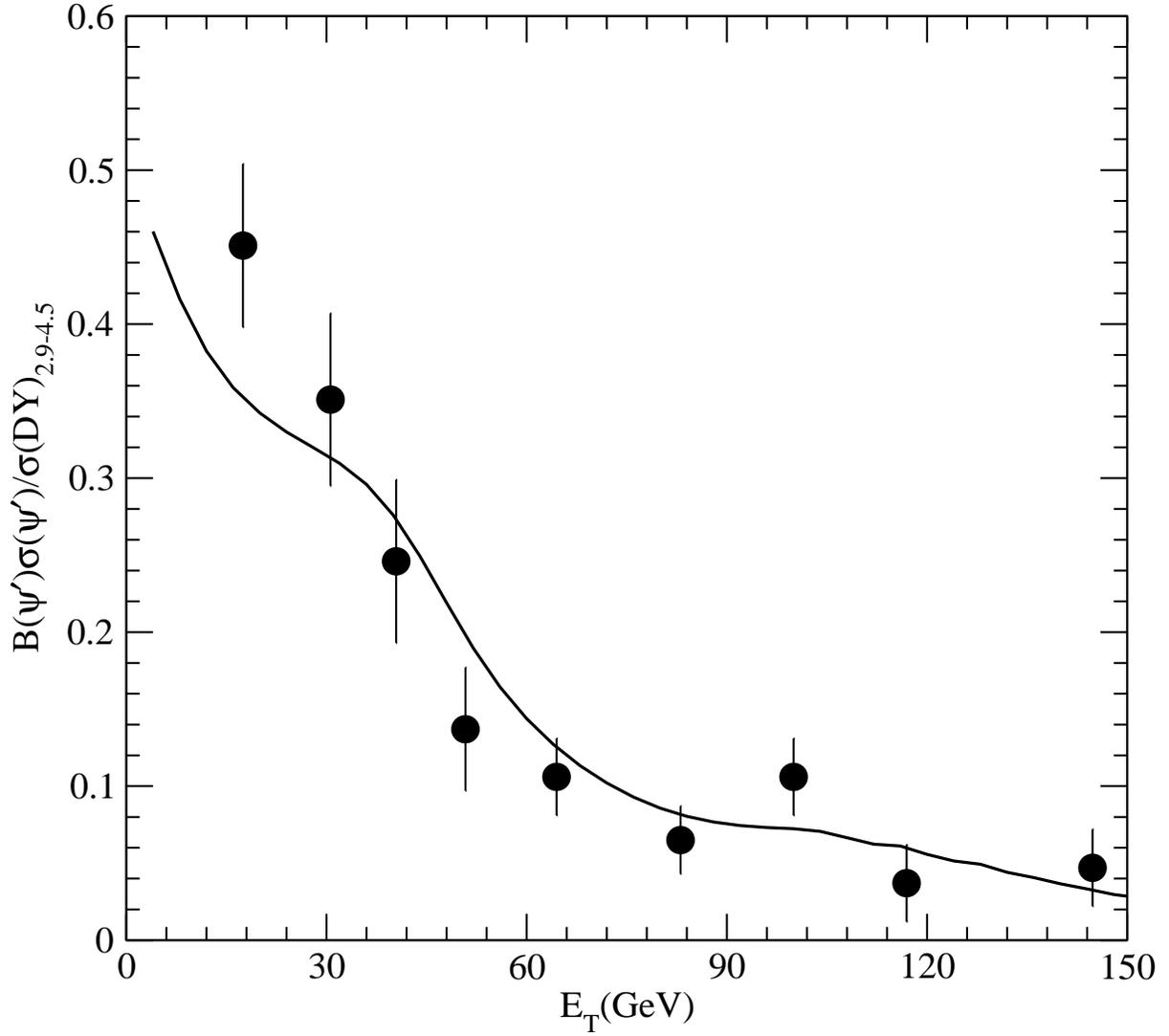}}
  \end{center}
\caption{
Solid curve obtained with $\sigma_{\rm Nabs} =4.2$ mb  is 
compared with the 1996  data of $\sigma (\psi') /\sigma (DY)$ ratio.
}
\label{fig4}
\end{figure}

\newpage
\begin{figure}[t]
  \begin{center}
    \leavevmode
    \parbox{\textwidth}
           {\psfig{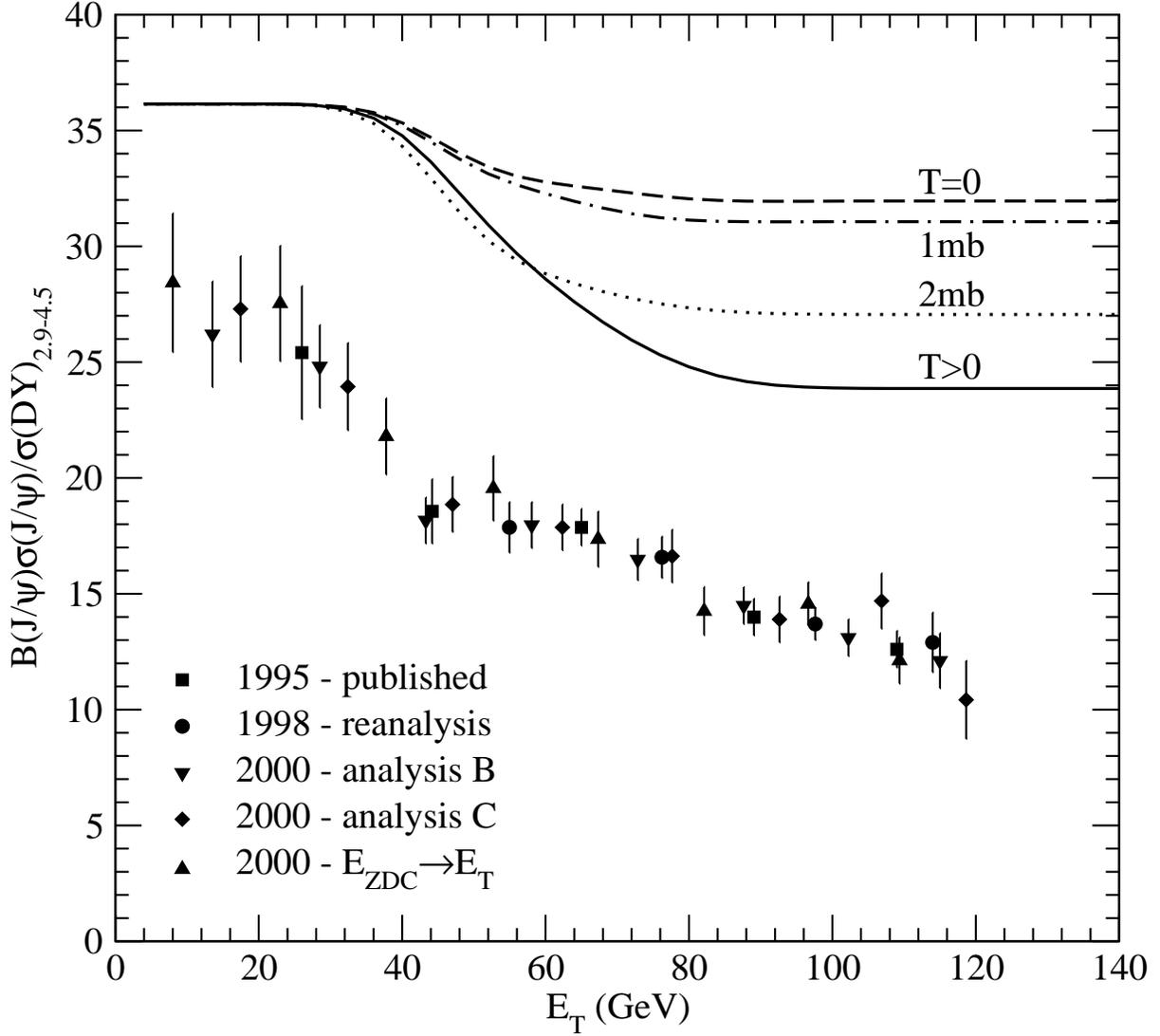}}
  \end{center}
\caption{
The same experimental data as Fig. 3. All curves have
$\sigma_{\rm Nabs} =0$ mb.
The solid, dashed, dotted, and dot-dashed curves are obtained by using the 
cross sections at $T>0$  and  
$T=0$, the two constant cross sections 
$\sigma_{\pi J/\psi}=\sigma_{\pi \chi_{cJ}}=\sigma_{\pi \psi'}=
2$ mb and 1 mb, respectively.
}
\label{fig5}
\end{figure}

\newpage
\begin{figure}[t]
  \begin{center}
    \leavevmode
    \parbox{\textwidth}
           {\psfig{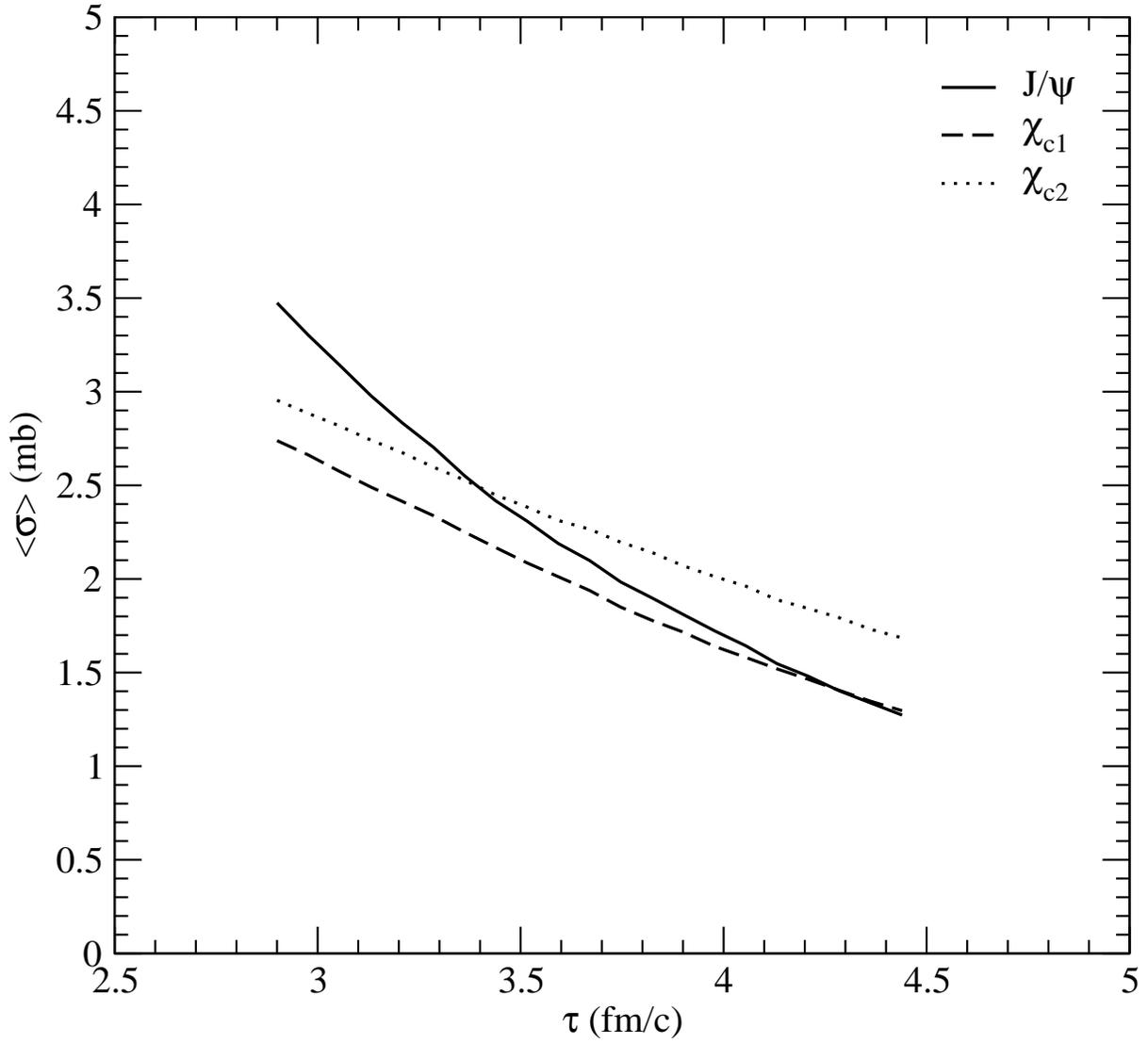}}
  \end{center}
\caption{
The solid, dashed, and dotted curves are 
$\langle \sigma_{\pi J/\psi} \rangle$, 
$\langle \sigma_{\pi \chi_{c1}} \rangle$, and
$\langle \sigma_{\pi \chi_{c2}} \rangle$, respectively.
}
\label{fig6}
\end{figure}

\newpage
\begin{figure}[t]
  \begin{center}
    \leavevmode
    \parbox{\textwidth}
           {\psfig{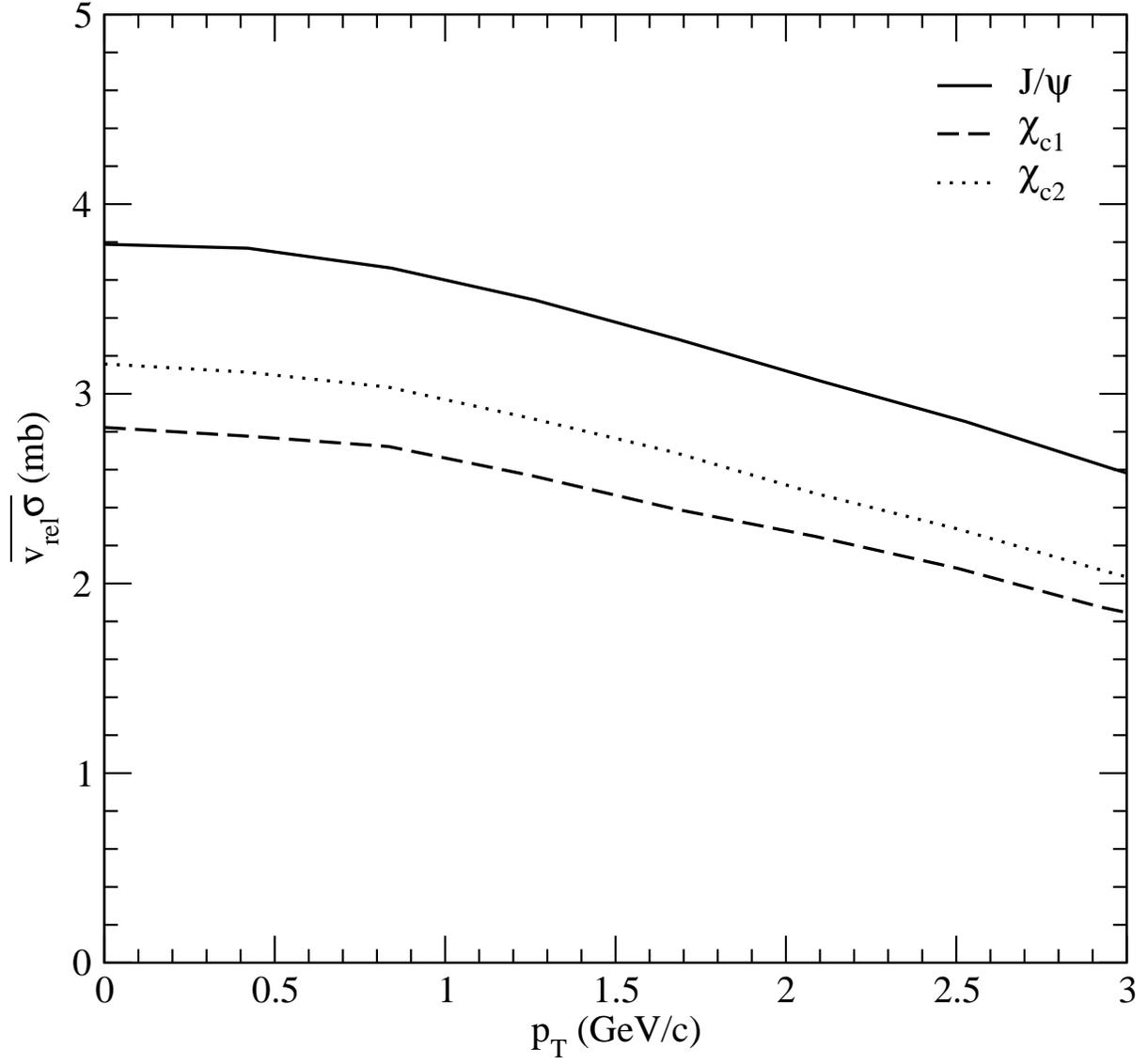}}
  \end{center}
\caption{
 The solid, dashed, and dotted  curves are 
$\overline {v_{\rm rel}\sigma_{\pi J/\psi}}$, 
$\overline {v_{\rm rel}\sigma_{\pi \chi_{c1}}}$, and
$\overline {v_{\rm rel}\sigma_{\pi \chi_{c2}}}$ at $x_F=0$, respectively.
}
\label{fig7}
\end{figure}

\newpage
\begin{figure}[t]
  \begin{center}
    \leavevmode
    \parbox{\textwidth}
           {\psfig{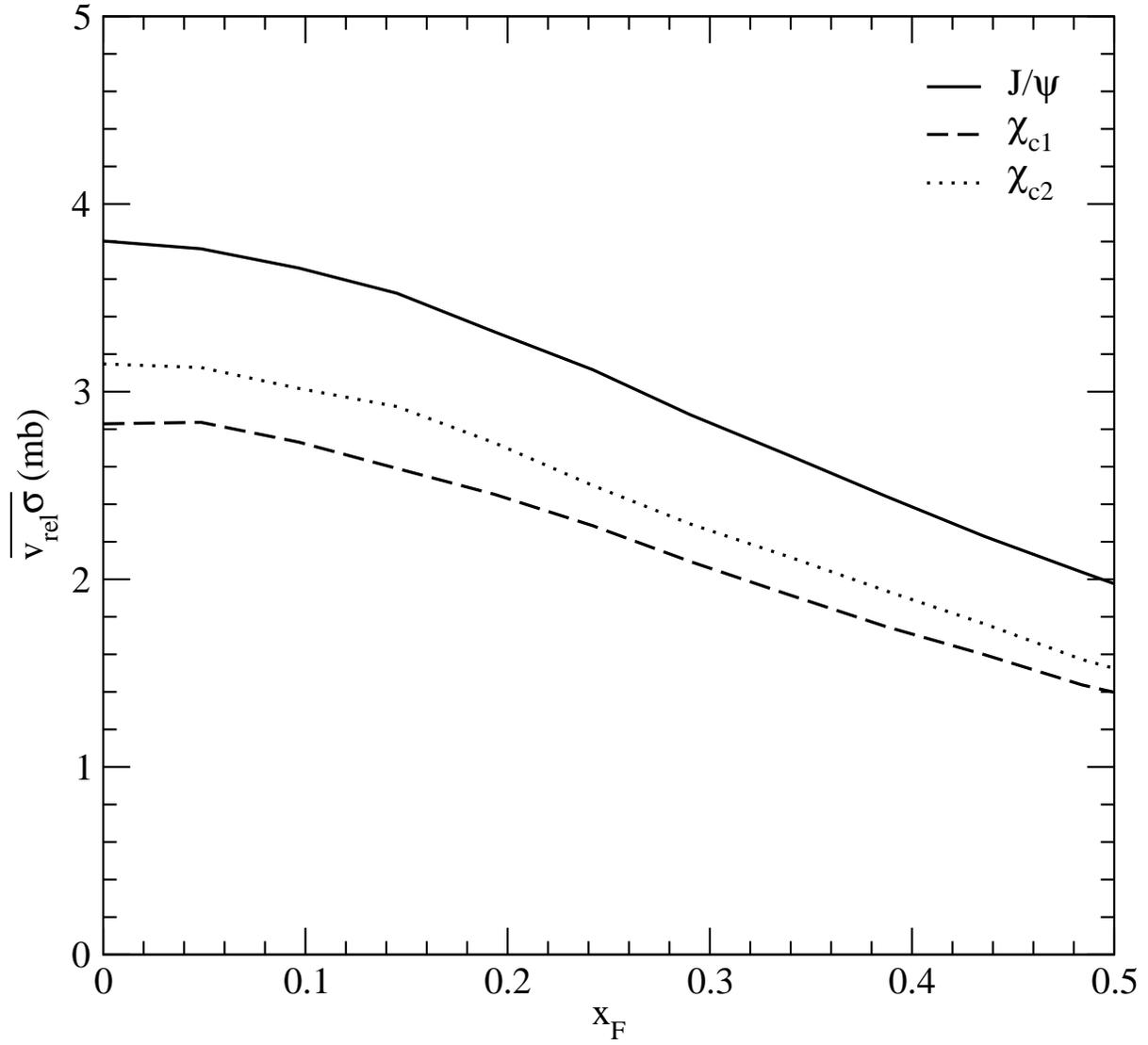}}
  \end{center}
\caption{
The same as Fig. 7 except for $x_F$ dependence at $p_T=0$. 
}
\label{fig8}
\end{figure}

\end{document}